\documentclass[useAMS,usenatbib]{mn2e}

\usepackage{graphicx}
\usepackage{amssymb}
\newcommand{\ion}[2]{\mbox{#1\,\textsc{#2}}}

\title[Neutral gas content of nine new galaxy candidates]{On the neutral gas content of nine new Milky Way satellite galaxy candidates}
\author[T.~Westmeier et al.]{\parbox{\textwidth}{T.~Westmeier$^{1}$\thanks{E-mail:
tobias.westmeier@uwa.edu.au (TW)}, L.~Staveley-Smith$^{1,2}$, M.~Calabretta$^{3}$, R.~Jurek$^{3}$, B.~S.~Koribalski$^{3}$, M.~Meyer$^{1,2}$, A.~Popping$^{1,2}$, O.~I.~Wong$^{1}$}\\ \\
$^{1}$ICRAR, M468, The University of Western Australia, 35~Stirling Highway, Crawley WA 6009, Australia\\
$^{2}$Australian Research Council, Centre of Excellence for All-sky Astrophysics (CAASTRO)\\
$^{3}$CSIRO Astronomy and Space Science, PO Box 76, Epping NSW 1710, Australia}
\begin{document}

\date{Accepted 1988 December 15. Received 1988 December 14; in original form 1988 October 11}

\pagerange{\pageref{firstpage}--\pageref{lastpage}} \pubyear{2002}

\maketitle

\label{firstpage}

\begin{abstract}
	We use a new, improved version of the \ion{H}{i} Parkes All-Sky Survey to search for \ion{H}{i}~emission from nine new, ultra-faint Milky Way satellite galaxy candidates recently discovered in data from the Dark Energy Survey. None of the candidates is detected in \ion{H}{i}, implying upper limits for their \ion{H}{i}~masses of typically several hundred to a few thousand solar masses. The resulting upper limits on $M_{\rm HI} / L_{\rm V}$ and $M_{\rm HI} / M_{\star}$ suggest that at least some of the new galaxy candidates are \ion{H}{i}~deficient. This finding is consistent with the general \ion{H}{i}~deficiency of satellite galaxies located within the Milky Way's virial radius and supports the hypothesis that gas is being removed from satellites by tidal and ram-pressure forces during perigalactic passages. In addition, some of the objects may be embedded in, and interacting with, the extended neutral and ionised gas filaments of the Magellanic Stream.
\end{abstract}


\begin{figure*}
	\includegraphics[width=\linewidth]{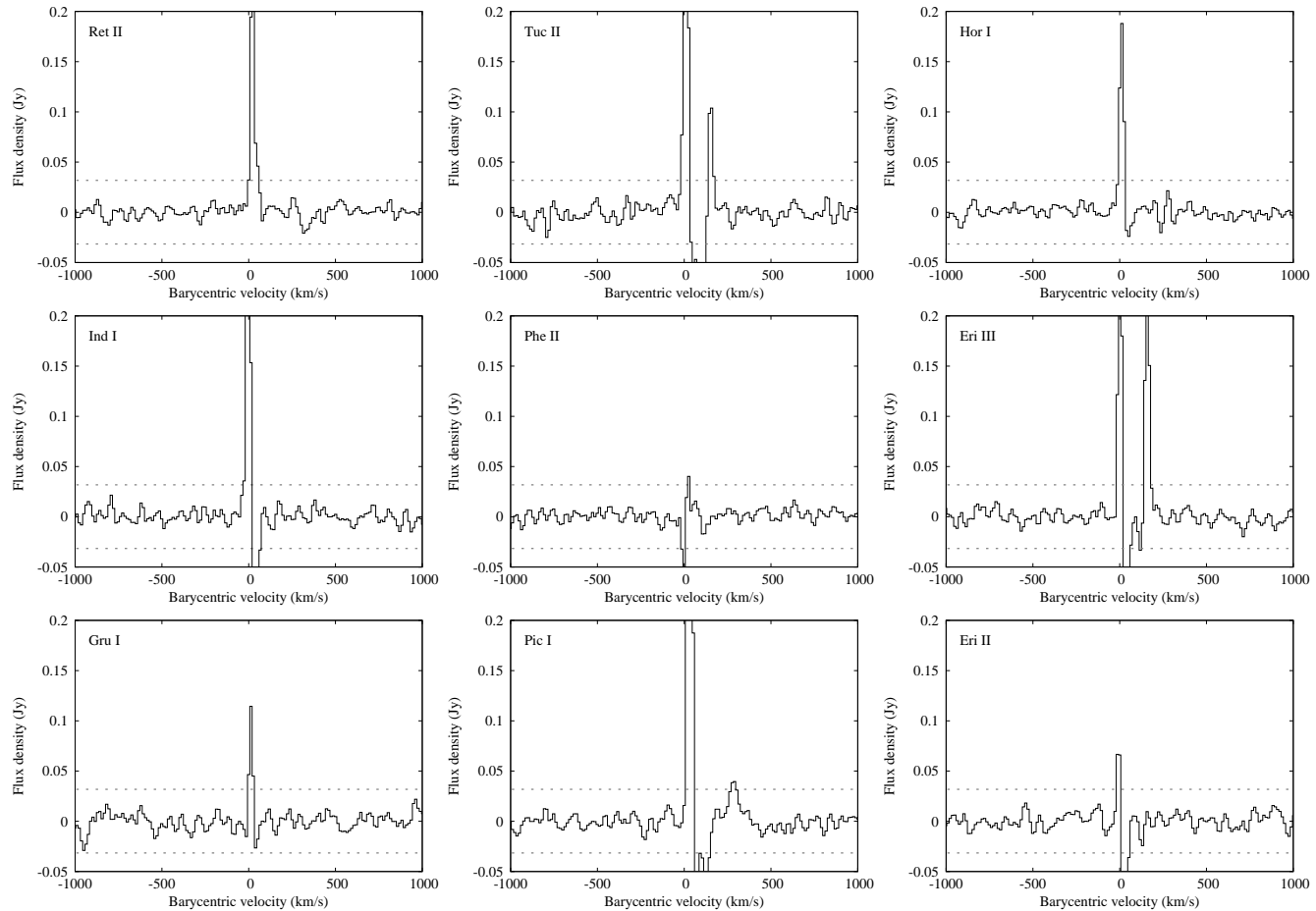}
	\caption{HIPASS~2 spectra in the velocity range of $-1000 < v_{\rm bar} < 1000~\mathrm{km \, s}^{-1}$ extracted at the positions of the new dwarf galaxy candidates. The dotted, grey lines show our mean $\pm 5 \langle \sigma \rangle$ sensitivity, where $\langle \sigma \rangle = 6.5~\mathrm{mJy}$. Strong positive and negative signals at $v_{\rm bar} \approx 0~\mathrm{km \, s}^{-1}$ are due to foreground emission from the Galactic disc. Other strong signals at positive velocities are likely due to the Magellanic Stream.}
	\label{fig_spectra}
\end{figure*}

\begin{figure*}
	\includegraphics[width=\linewidth]{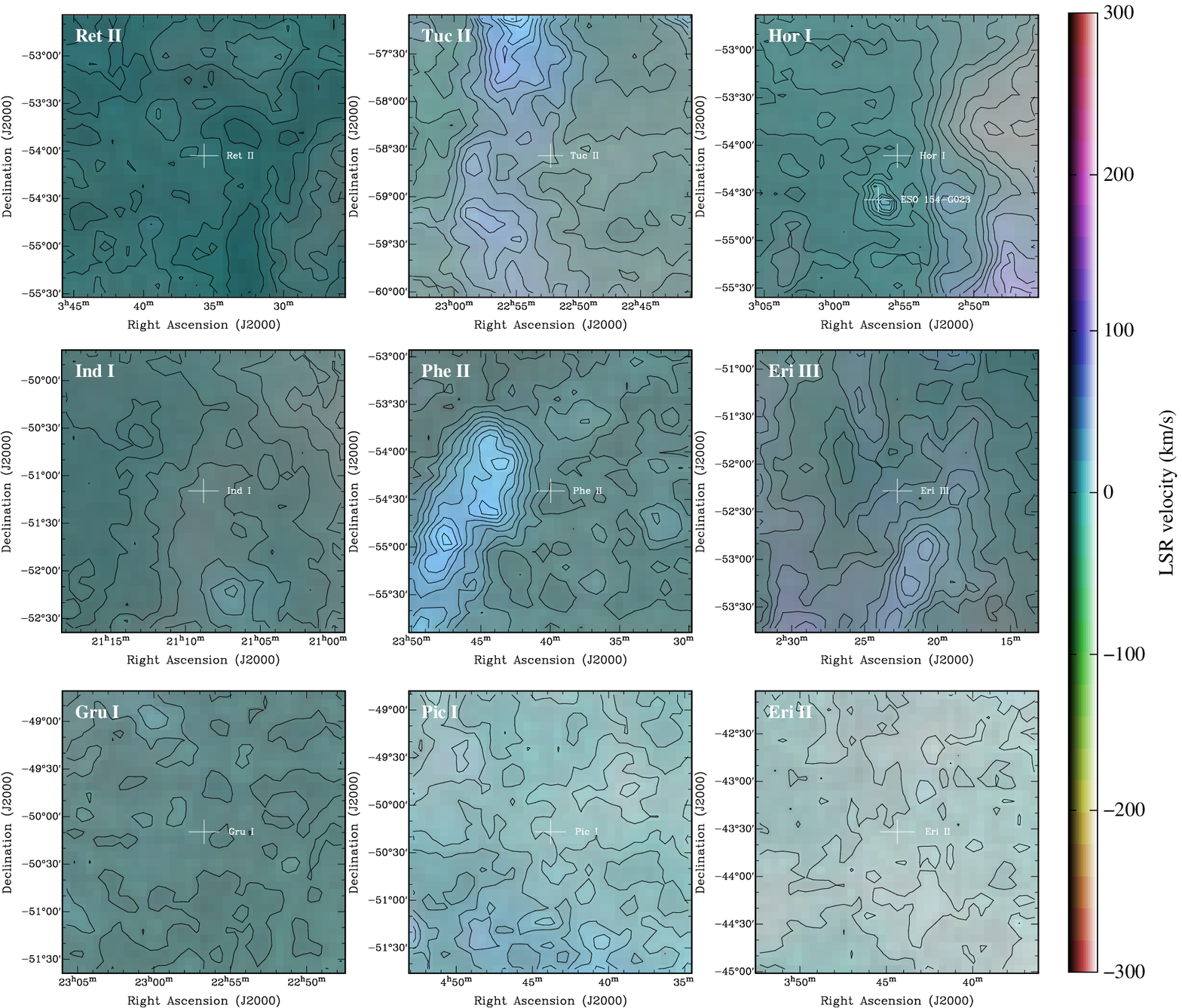}
	\caption{Integrated GASS \ion{H}{i}~maps of the field around each of the galaxy candidates in the velocity range of $-300 < v_{\rm LSR} < 300~\mathrm{km \, s}^{-1}$. In the colour images, hue represents velocity as indicated by the colour bar, while lightness represents the sum of logarithmic flux densities, $\sum \log[(S / \mathrm{Jy}) + 1]$, in the range of $1$ (black) to $5$ (white), with contours drawn in intervals of $0.1$. The positions of all galaxy candidates and the background galaxy ESO~154$-$G023 are marked with a white cross. Extended greenish hues indicate foreground emission from the Galactic disc, while blue and purple hues generally correspond to \ion{H}{i}~emission from the Magellanic Stream.}
	\label{fig_maps}
\end{figure*}

\section{Introduction}

Observations of neutral hydrogen (\ion{H}{i}) in dwarf galaxies throughout the Local Group have revealed that satellite galaxies of the Milky Way and the Andromeda Galaxy are generally \ion{H}{i}-deficient \citep{Young1999,Young2000,Grcevich2009}, in particular those located within the virial radius of the Milky Way of $R_{\rm vir} \approx 300~\mathrm{kpc}$ \citep{Spekkens2014}. This \ion{H}{i}~deficiency has generally been interpreted as being the result of increased tidal and ram-pressure forces in the vicinity of the most massive members of the Local Group, although photoionisation and heating of the gas may also play a role \citep{Efstathiou1992}. If correct, this hypothesis would have significant consequences for the star formation history of Milky Way satellites \citep{Bahe2015} and the interpretation of the so-called missing satellites problem \citep{Nickerson2011}.

Recently, \citet{Bechtol2015} and \citet{Koposov2015} reported the discovery of a total of nine new, ultra-faint Milky Way satellite galaxy candidates in data from the Dark Energy Survey \citep{Abbott2005}. These galaxy candidates are located in the vicinity of the Magellanic Clouds and parts of the Magellanic Stream on the sky at distances between about $30$ and $360~\mathrm{kpc}$ from the Milky Way. Their locations suggest that some of them could be physically associated with the Magellanic system and form a population of Magellanic satellites. Three of the satellite candidates lie beyond $100~\mathrm{kpc}$, raising the question of whether they may still contain some neutral gas. This could in particular be the case for the most distant object, Eri~II, which possibly contains a younger stellar population \citep{Koposov2015}.

Here, we describe our search for potential 21-cm line emission of neutral hydrogen associated with the nine satellite candidates, using a new and improved version of the \ion{H}{i} Parkes All-Sky Survey. We first introduce the data sets used for this work in Section~\ref{sect_data} before describing the results of our search in Section~\ref{sect_results}. This is followed by a discussion of our findings in Section~\ref{sect_discussion} and finally a brief summary of our main results and conclusions in Section~\ref{sect_summary}.

\begin{figure*}
	\includegraphics[width=0.8\linewidth]{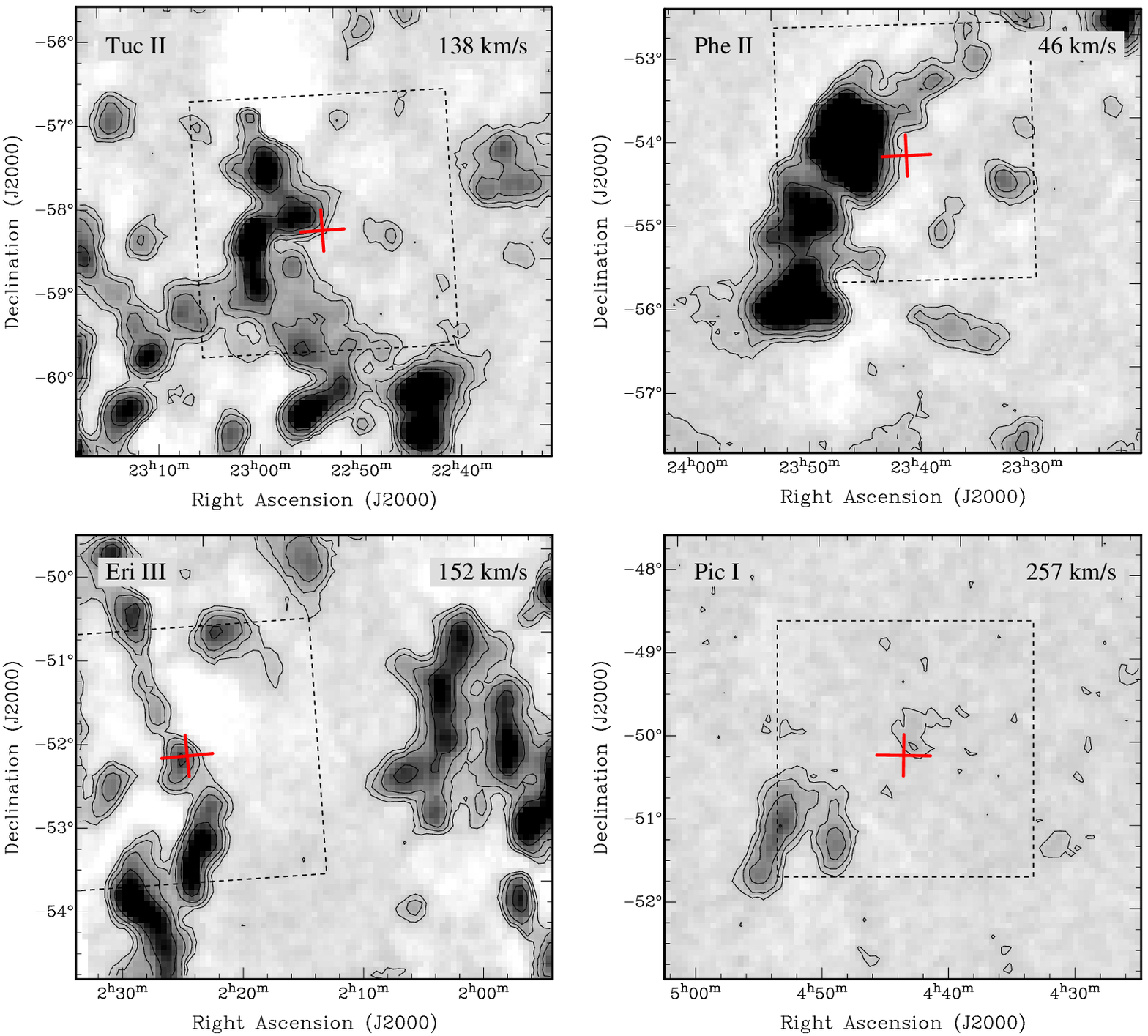}
	\caption{Individual channel maps from HIPASS~2 of a larger area around the four candidates with \ion{H}{i}~emission lines in their spectrum at positive velocities (Tuc~II, Phe~II, Eri~III and Pic~I). The velocities correspond to the peak of the respective \ion{H}{i}~emission line at positive velocities in Fig.~\ref{fig_spectra}. Contour lines are drawn at levels of $0.02$, $0.05$, $0.1$, $0.2$, $0.5$, $1$ and $2~\mathrm{Jy}$. The position of each candidate is marked with a red cross, while the black square outlines the approximate field of view of the corresponding map in Fig.~\ref{fig_maps} (note that the coordinate system may not be aligned with the axes of the plot). In each case, the \ion{H}{i}~signal stems from extended, clumpy filaments most likely associated with the Magellanic Stream.}
	\label{fig_chanmaps}
\end{figure*}

\section{Data}
\label{sect_data}

The results presented in this paper are based on HIPASS~2, a new and improved version of the \ion{H}{i}~Parkes All-Sky Survey (HIPASS; \citealt{Barnes2001}). The new version uses the original HIPASS raw data, but makes use of several new data reduction techniques that significantly reduce the number of artefacts in the data and slightly decrease noise levels in the final data cubes. Details of most of the improvements are discussed in a recent paper by \citet{Calabretta2014} describing the new $1.4~\mathrm{GHz}$ continuum map derived from HIPASS data. Another paper specifically describing the new HIPASS~2 spectral-line data is in preparation. The velocity resolution of HIPASS~2 is $26.4~\mathrm{km \, s}^{-1}$ after application of a Hann filter along the spectral axis.

In addition to HIPASS~2, we also made use of the latest data release \citep{Kalberla2015} of the Galactic All-Sky Survey (GASS; \citealt{McClure-Griffiths2009,Kalberla2010}). While both HIPASS~2 and GASS are \ion{H}{i}~surveys observed with the multi-beam receiver at the Parkes radio telescope, there are a few significant differences between the two surveys that make them complementary to some degree. HIPASS~2 was designed to be an extragalactic survey covering a redshift range of $-1280 \lesssim \mathrm{c}z \lesssim +12\,700~\mathrm{km \, s}^{-1}$ and suffers from strong data reduction artefacts in frequency channels dominated by extended \ion{H}{i}~emission from the Galactic disc and the Magellanic system. GASS, in contrast, is a Galactic \ion{H}{i}~survey that covers a redshift range of $-450 \lesssim \mathrm{c}z \lesssim +450~\mathrm{km \, s}^{-1}$, has a much higher velocity resolution of about $0.8~\mathrm{km \, s}^{-1}$ and fully recovers all diffuse flux without introducing strong artefacts. This would make GASS the more suitable survey for our search for \ion{H}{i}~emission from the new Milky Way satellite galaxy candidates. However, HIPASS~2 is more sensitive than GASS by almost a factor of~4, with a total integration time per pointing of $450~\mathrm{s}$ for HIPASS~2 as compared to $30~\mathrm{s}$ for GASS. Hence, we use the HIPASS~2 data in our search for \ion{H}{i}~emission from the new satellite candidates and for the purpose of establishing detection thresholds, while GASS data are used for detailed imaging of their environment.

\section{Results}
\label{sect_results}

The basic observational parameters of the nine galaxy candidates are summarised in Table~\ref{tab_parameters}. All optically derived parameters were extracted from \citet{Bechtol2015} and \citet{Koposov2015} and, where possible, the weighted mean of the two measurements was taken. To determine the distance of each satellite candidate from the Milky Way, the measured heliocentric distances were converted to Galactocentric distances by subtracting the heliocentric space vector of the Galactic centre, $\vec{R⃗}_{\rm GC}$, from that of the satellite candidate, assuming a distance to the Galactic centre of $|\vec{R⃗}_{\rm GC}| = 8~\mathrm{kpc}$ \citep{Eisenhauer2003,Do2013}.

In order to search for a possible neutral gas component associated with the galaxy candidates, we extracted the \ion{H}{i}~spectrum at the position of each of the candidates from the HIPASS~2 data. The assumption was made that any \ion{H}{i}~component would be centred on the optical position of the galaxy candidate and be spatially unresolved by the $15~\mathrm{arcmin}$ Parkes beam (equivalent to a physical resolution of $440~\mathrm{pc}$ at a distance of $100~\mathrm{kpc}$). We then used the \textsc{Miriad} task \textsc{mbspect} with the parameter \texttt{yaxis=point} to extract the \ion{H}{i}~spectrum at the position of each source. This will generate a beam-weighted spectrum at the given position under the assumption that the object is a point source. The resulting spectra are displayed in Fig.~\ref{fig_spectra}.

We also generated integrated flux density maps of the region around each source from publicly available GASS data. In order to capture both the flux density and radial velocity of the \ion{H}{i}~gas, we synthesised false-colour maps in which lightness reflects the flux density of the emission, while hue encodes the velocity of the gas \citep{Rector2007}. The resulting integrated flux density maps are presented in Fig.~\ref{fig_maps}. They will be helpful in imaging the gaseous structures in each region and identifying any potential spectral components present in the HIPASS~2 spectrum.

\section{Discussion}
\label{sect_discussion}

Each spectrum in Fig.~\ref{fig_spectra} shows strong positive or negative components at $v_{\rm bar} \approx 0~\mathrm{km \, s}^{-1}$. These are due to foreground emission from the Galactic disc. Negative signals are simply due to extended Galactic emission getting subtracted during the bandpass calibration of HIPASS which -- being primarily an extragalactic survey -- had been optimised for compact sources. A few spectra show an additional narrow emission line at positive velocities in the range of about $50$ to $300~\mathrm{km \, s}^{-1}$, most notably those of Tuc~II, Phe~II, Eri~III, and Pic~I. Closer inspection of the maps in Fig.~\ref{fig_maps} reveals that in all of these cases there are extended blue structures across large parts of the map. This emission is likely associated with the Magellanic Stream and not part of any of the satellite galaxy candidates. To confirm this, we plot in Fig.~\ref{fig_chanmaps} individual \ion{H}{i} channel maps from HIPASS~2 of a larger region around each of those four candidates. The channel maps reveal the presence of extended, clumpy gas filaments at the velocities of the \ion{H}{i}~lines in Fig.~\ref{fig_spectra}. Due to their large size and coherent spatial and kinematic structure, these filaments are almost certainly part of the Magellanic Stream and unlikely to be connected to any of the galaxy candidates.

\begin{table*}
	\caption{Physical parameters of the nine satellite galaxy candidates. The columns denote the name of the candidate, Galactic longitude and latitude, heliocentric distance, Galactocentric distance, rms noise level of the extracted spectrum, $5 \times \sigma_{\rm rms}$~upper limit of the \ion{H}{i}~mass across $\Delta v = 26.4~\mathrm{km \, s}^{-1}$, stellar mass, upper limit of \ion{H}{i} over stellar mass, V-band luminosity, and upper limit of \ion{H}{i}~mass over V-band luminosity. Optical parameters ($d_{\rm hel}$, $M_{\star}$ and $L_{\rm V}$) have been extracted from \citet{Bechtol2015} and \citet{Koposov2015} and, where possible, represent the weighted mean of the two.}
	\label{tab_parameters}
	\begin{tabular}{lrrrrrrrrrr}
		\hline
		Name    & $l$       & $b$       & $d_{\rm hel}$     & $d_{\rm gal}$ & $\sigma_{\rm rms}$ & $\log(M_{\rm HI}$ & $\log(M_{\star}$       & $\log(M_{\rm HI}$ & $\log(L_{\rm V}$ & $\log[(M_{\rm HI} / L_{\rm V})$ \\
		        & ($\degr$) & ($\degr$) & (kpc)             & (kpc)         & (mJy)              & $/M_{\sun})$      & $/M_{\sun})$           & $ / M_{\star})$   & $/L_{\sun})$     & $/(M_{\sun} / L_{\sun})]$  \\
		\hline
		Ret~II  & $266.299$ & $-49.738$ & $ 31^{ +2}_{ -2}$ &          $32$ &              $5.2$ &          $< 2.19$ & $3.41^{+0.03}_{-0.03}$ &         $< -1.22$ &  $3.19 \pm 0.03$ & $< -1.00$ \\
		Tuc~II  & $328.039$ & $-52.354$ & $ 58^{ +4}_{ -4}$ &          $54$ &              $6.7$ &          $< 2.84$ & $3.48^{+0.52}_{-0.18}$ &         $< -0.64$ &  $3.46 \pm 0.04$ & $< -0.62$ \\
		Hor~I   & $271.382$ & $-54.744$ & $ 83^{ +6}_{ -5}$ &          $84$ &              $5.4$ &          $< 3.07$ & $3.38^{+0.20}_{-0.15}$ &         $< -0.31$ &  $3.30 \pm 0.04$ & $< -0.23$ \\
		Ind~I   & $347.165$ & $-42.069$ & $ 85^{ +6}_{ -5}$ &          $79$ &              $7.4$ &          $< 3.22$ & $2.90^{+0.18}_{-0.30}$ &         $< +0.31$ &  $3.26 \pm 0.07$ & $< -0.04$ \\
		Phe~II  & $323.688$ & $-59.744$ & $ 89^{ +6}_{ -6}$ &          $86$ &              $6.4$ &          $< 3.20$ & $3.45^{+0.15}_{-0.12}$ &         $< -0.25$ &  $3.12 \pm 0.07$ & $< +0.08$ \\
		Eri~III & $274.953$ & $-59.601$ & $ 91^{ +6}_{ -6}$ &          $91$ &              $6.8$ &          $< 3.25$ & $2.95^{+0.30}_{-0.65}$ &         $< +0.29$ &  $2.76 \pm 0.11$ & $< +0.48$ \\
		Gru~I   & $338.682$ & $-58.248$ & $120^{+12}_{-11}$ &         $117$ &              $6.7$ &          $< 3.48$ &                     -- &                -- &  $3.29 \pm 0.12$ & $< +0.18$ \\
		Pic~I   & $257.295$ & $-40.644$ & $120^{ +8}_{ -7}$ &         $122$ &              $7.0$ &          $< 3.50$ & $3.45^{+0.44}_{-0.41}$ &         $< +0.05$ &  $3.26 \pm 0.10$ & $< +0.24$ \\
		Eri~II  & $249.776$ & $-51.644$ & $356^{+24}_{-22}$ &         $357$ &              $6.6$ &          $< 4.41$ & $4.92^{+0.08}_{-0.08}$ &         $< -0.51$ &  $4.73 \pm 0.03$ & $< -0.32$ \\
		\hline
	\end{tabular}
\end{table*}

This leads us to conclude that no \ion{H}{i}~emission was detected in HIPASS~2 from any of the nine galaxy candidates, and we will only be able to provide upper limits on their \ion{H}{i}~masses. In principle it would be possible for the velocities of some of the galaxy candidates to overlap with those of the Galactic or Magellanic emission, which would make them difficult to detect. This is a general problem for all local \ion{H}{i}~sources whose radial velocities are expected to be close to $0~\mathrm{km \, s}^{-1}$. Under certain conditions it would still be possible to discern such sources even if their spectral lines perfectly blended in with the Galactic foreground. This would require the foreground emission to be spatially extended and smooth, while the background source would need to be sufficiently compact. An example of this scenario is ESO~154$-$G023, a galaxy located close to one of the galaxy candidates, Hor~I, and visible in the top-right panel of Fig.~\ref{fig_maps}. In the GASS data its \ion{H}{i}~emission covers the LSR velocity range of approximately $-100$ to $+150~\mathrm{km \, s}^{-1}$, yet it is bright and compact enough to stand out against the Galactic foreground emission despite the large overlap in velocity. Curiously, ESO~154$-$G023 is listed in the HIPASS Bright Galaxy Catalogue \citep{Koribalski2004} as having a radial velocity of $v_{\rm bar} = +574~\mathrm{km \, s}^{-1}$, indicating that the emission in GASS at $v_{\rm LSR} \approx {-100}$ to $+150~\mathrm{km \, s}^{-1}$ is an artefact in the form of a ghost image produced by either the multi-beam receiver and correlator system at Parkes or the GASS data reduction pipeline. Given the positional coincidence between the \ion{H}{i}~signal and ESO~154$-$G023 as well as the large separation of more than half a degree from the position of Hor~I, it would appear unlikely that the emission at that position is associated in any way with Hor~I.

\begin{figure*}
	\includegraphics[width=0.49\linewidth]{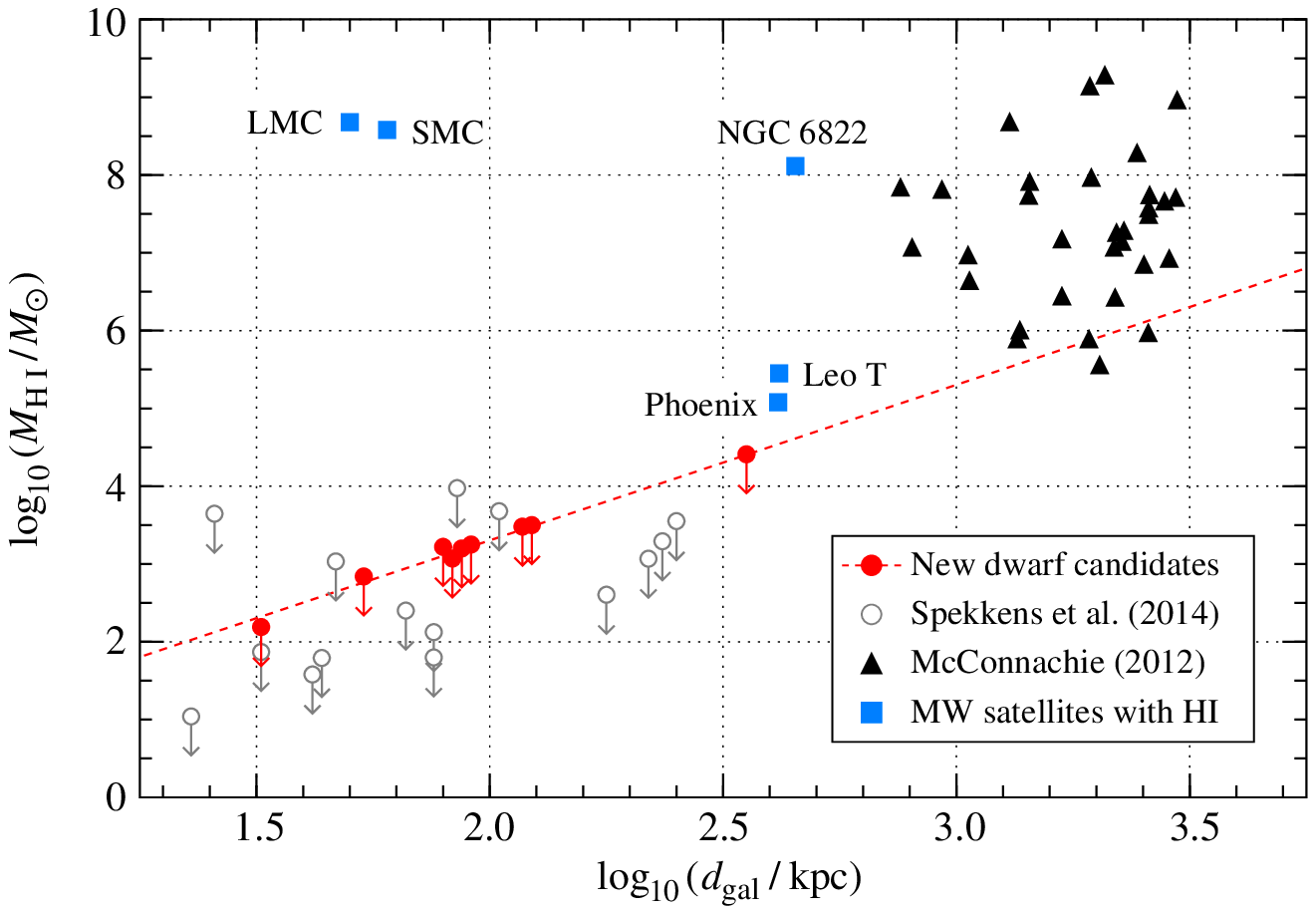}
	\hfill
	\includegraphics[width=0.49\linewidth]{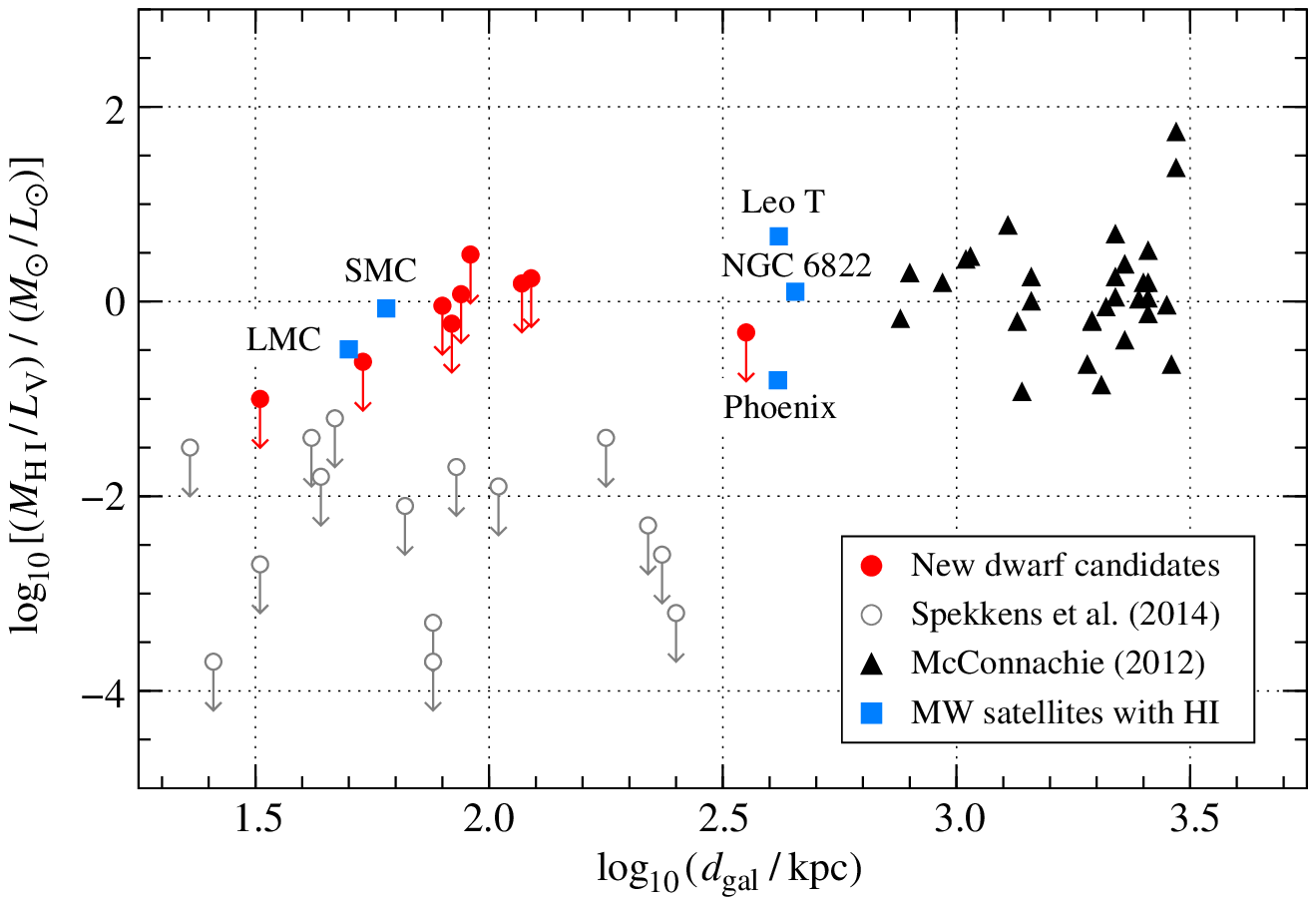}
	\caption{Upper limits on the \ion{H}{i}~mass (left) and \ion{H}{i}~mass over V-band luminosity (right) for the nine dwarf galaxy candidates studied here (red, filled circles) in comparison to the upper limits for the Milky Way satellites studied by \citet{Spekkens2014} (grey, open circles) and the \ion{H}{i}~detections of Local Group and neighbouring galaxies listed by \citet{McConnachie2012} (black, filled triangles), plotted against Galactocentric distance, $d_{\rm gal}$. The dashed, red line represents our mean $5 \sigma$ detection threshold over $\Delta v = 26.4~\mathrm{km \, s}^{-1}$. Galactic satellites with unambiguous \ion{H}{i} detections are plotted as the blue squares for comparison.}
	\label{fig_m-l}
\end{figure*}

The upper \ion{H}{i}~mass limits resulting from the non-detections in HIPASS~2 are presented in Table~\ref{tab_parameters} and the left-hand panel of Fig.~\ref{fig_m-l}, assuming a conservative $5 \sigma$ detection threshold across a velocity width of $26.4~\mathrm{km \, s}^{-1}$. They are typically of the order of a few thousand times the mass of the Sun, thus being among the lowest \ion{H}{i}~mass limits ever measured in galaxies. In order to assess the significance of these non-detections in terms of \ion{H}{i}~deficiency, we plot the derived upper limits of $M_{\rm HI} / L_{\rm V}$, where $L_{\rm V}$ is the optical V--band luminosity of the galaxy candidates, as a function of Galactocentric distance in the right-hand panel of Fig.~\ref{fig_m-l}. For comparison we also show the \ion{H}{i} non-detections of known Milky Way satellite galaxies by \citet{Spekkens2014} as well as the \ion{H}{i}~detections of galaxies in the Local Group and neighbouring regions that are not classified as Milky Way or Andromeda satellites by \citet{McConnachie2012}. Most galaxy candidates have upper limits on $\log[(M_{\rm HI} / L_{\rm V}) / (M_{\sun} / L_{\sun})]$ that are near the median of $+0.015$ of the Local Group and neighbouring detections listed by \citet{McConnachie2012}. Two candidates have substantially lower values, the lowest being Ret~II with an upper limit of $-1.0$, an order of magnitude below the median of the detections.

Similar and generally even stricter upper limits are also obtained for the \ion{H}{i}-to-stellar mass ratio, $\log(M_{\rm HI} / M_{\star})$, for those eight candidates for which stellar mass measurements have been published by \citet{Bechtol2015}, with Ret~II again having the strictest limit of $-1.2$. Our results suggest that at least some of the new satellite candidates are \ion{H}{i}-deficient in comparison to Local Group and neighbouring galaxies detected in \ion{H}{i}~emission as reported by \citet{McConnachie2012}.

For comparison, we also include in Fig.~\ref{fig_m-l} the measurements for likely Milky Way satellites with unambiguous \ion{H}{i} detections, including the LMC \citep{Staveley-Smith2003}, the SMC \citep{Stanimirovic1999}, Phoenix\footnote{The \ion{H}{i}~cloud at a barycentric velocity of $\mathrm{c}z \approx {-23}~\mathrm{km \, s}^{-1}$ partially overlaps with the stellar component of Phoenix and is also consistent with the optical velocity measurement by \citet{Irwin2002}.} \citep{Young1997,Young2007}, NGC~6822 \citep{deBlok2000} and Leo~T \citep{Simon2007,Ryan-Weber2008}. Their values of $M_{\rm HI} / L_{\rm V}$ are generally consistent with those of the Local Group and neighbouring galaxies from \citet{McConnachie2012}. Note that there is strong evidence for \ion{H}{i} gas currently being removed from the Phoenix dwarf galaxy through either ram-pressure stripping by the intergalactic medium or ejection by winds related to supernovae and star formation activity \citep{StGermain1999,Gallart2001}. This would explain its comparatively low \ion{H}{i}~content of $\log[(M_{\rm HI} / L_{\rm V}) / (M_{\sun} / L_{\sun})] \approx {-0.8}$.

Under the general assumption that the new galaxy candidates are indeed galaxies (as opposed to star clusters; see the discussion in \citealt{Koposov2015}), our upper limits on $M_{\rm HI} / L_{\rm V}$ and $M_{\rm HI} / M_{\star}$ are consistent with previous findings that dwarf galaxies within the virial radius of the Milky Way of $R_{\rm vir} \approx 300~\mathrm{kpc}$ are extremely \ion{H}{i} deficient \citep{Young1999,Young2000,Grcevich2009,Spekkens2014}. \citet{Spekkens2014} conclude that these galaxies have likely had their gas stripped due to a combination of tidal and ram-pressure interaction in the past. Such interactions are unlikely to occur continuously along the entire orbit of the satellite, but instead will be more frequent and stronger near perigalacticon or whenever the satellite crosses the Galactic plane. With the exception of Eri~II, all of our targets are well within the virial radius of the Milky Way, and our \ion{H}{i} non-detections fit into this picture.

While our upper \ion{H}{i}~mass limits are similar to those measured by \citet{Spekkens2014} for known Milky Way satellites, the resulting upper limits on $M_{\rm HI} / L_{\rm V}$ are typically several orders of magnitude larger. This difference is largely due to the extremely low optical luminosities of the new satellite galaxy candidates of typically only a few thousand $L_{\sun}$, which is several orders of magnitude below the luminosities of most of the galaxies studied by \citet{Spekkens2014}. Unfortunately, pushing our upper limits on $M_{\rm HI} / L_{\rm V}$ down into the regime probed by \citet{Spekkens2014} would require prohibitive integration times. For example, an improvement of our measurement by just one order of magnitude would require one hundred times the integration time of HIPASS~2, equivalent to 12.5~h of continuous integration per source on the 64-m Parkes radio telescope. While such a time request might still be acceptable for a small sample of targets, the presence of low-level radio frequency interference and systematic effects in the receiver/correlator system could potentially limit the telescope's sensitivity at such extreme integration times. Hence, we have reached the limit of what can be practically achieved with existing telescopes, and future instruments like the Square Kilometre Array will be necessary to probe the regime of ultra-faint dwarf galaxies.

If the new galaxy candidates are indeed associated with the Magellanic Clouds, as suggested by \citet{Koposov2015}, gas may have been lost during previous perigalactic passages of the Magellanic system. This scenario is supported by new orbital simulations of the candidates by \citet{Yozin2015}, although it depends on assumptions made about the past orbital history of the Magellanic Clouds (see the discussion in \citealt{Kallivayalil2013}). Early gas loss is suggested by the generally old stellar populations ($\tau \simeq 10~\mathrm{Ga}$) as determined by \citet{Bechtol2015} for the four most significant galaxy candidate detections. Interestingly, \citet{Koposov2015} report possible evidence for more recent star formation in the most distant galaxy candidate, Eri~II, consistent with its current location just outside the virial radius of the Milky Way. Their finding suggests that some \ion{H}{i}~gas may still be present in Eri~II, although below the upper limit of about $2.5 \times 10^{4}~M_{\odot}$ reported here.

The non-detection of \ion{H}{i}~emission from the nine satellite galaxy candidates could also be the result of their close proximity to the Magellanic Clouds. In fact, the positions of several candidates on the sky overlap with \ion{H}{i}~emission from the Magellanic Stream (see figure~22 of \citealt{Koposov2015}), and their distances from the Milky Way are similar to those expected for the Stream in that direction (\citealt{Diaz2012}), suggesting that some of the galaxy candidates could actually be physically embedded in the neutral \citep{Putman2003,Bruens2005} and ionised \citep{Weiner1996,Sembach2003} gas of the Magellanic Stream. This scenario is even more credible in view of the findings by \citet{Fox2014} that the Stream contains three times as much ionised gas as neutral gas and is substantially more extended than indicated by the \ion{H}{i} component alone. In this case, the expected strong ram-pressure interaction with the gas of the Stream would almost certainly have removed any remaining gas component from the affected galaxy candidates.

\section{Summary}
\label{sect_summary}

We extracted \ion{H}{i}~spectra from HIPASS~2 at the positions of nine new Milky Way satellite galaxy candidates recently reported by \citet{Bechtol2015} and \citet{Koposov2015} to search for a potential neutral gas component associated with these candidates. Eight of the nine candidates are positioned well within the virial radius of the Milky Way, with one, Eri~II, located just outside. Our main results and conclusions are as follows:
\begin{enumerate}
	\item We do not detect any \ion{H}{i}~signal likely to be associated with any of the nine galaxy candidates, including the most distant object, Eri~II. The resulting upper \ion{H}{i}~mass limits are typically in the range of several hundred to a few thousand solar masses, with a slightly higher limit of about $2.5 \times 10^{4}~M_{\odot}$ for Eri~II.
	\item The resulting upper limits on $M_{\rm HI} / L_{\rm V}$ are generally near or below the median value of \ion{H}{i}~detections in the Local Group and neighbouring regions as reported by \citet{McConnachie2012}, suggesting that at least some of the new galaxy candidates are \ion{H}{i}-deficient compared to Local Group galaxies residing outside the virial radius of the Milky Way and the Andromeda galaxy. Similar conclusions are derived from upper limits on $M_{\rm HI} / M_{\star}$.
	\item This result is consistent with the hypothesis that satellite galaxies located within the virial radius of the Milky Way are generally \ion{H}{i}-deficient as a result of gas stripping due to increased tidal and ram-pressure forces occurring near the perigalacticon of their orbit. Furthermore, several candidates may actually be embedded in, and thus interacting with, the neutral and ionised gas filaments of the Magellanic Stream.
	\item Pushing our current limits on $M_{\rm HI} / L_{\rm V}$ for such ultra-faint dwarf galaxies down into a more interesting regime will be difficult with existing instruments due to the prohibitive integration times required.
\end{enumerate}

\section*{Acknowledgments}

The authors wish to thank K.~Bekki for insightful discussions about the possible interaction history of the galaxy candidates. The Parkes radio telescope is part of the Australia Telescope National Facility which is funded by the Commonwealth of Australia for operation as a National Facility managed by CSIRO. Parts of this research were conducted by the Australian Research Council Centre of Excellence for All-sky Astrophysics (CAASTRO) through project number CE110001020.

\bsp

\label{lastpage}

\end{document}